# Title: Abiotic Ozone and Oxygen in Atmospheres Similar to Prebiotic Earth

**Short Title:** Oxygen and Ozone on Lifeless Planets

**Authors:** Shawn D. Domagal-Goldman[1,2]*, Antígona Segura[2,3], Mark W. Claire[2,4], Tyler D. Robinson[2,5], Victoria S. Meadows[2,6]

**Affiliations:**

[1]Planetary Environments Laboratory, NASA Goddard Space Flight Center, 8800 Greenbelt Road, Greenbelt, MD 20771, USA

[2]NASA Astrobiology Institute – Virtual Planetary Laboratory

[3]Instituto de Ciencias Nucleares, Universidad Nacional Autónoma de México, Circuito Interior S/N, C.U. A.P. 70-543, Mexico , D.F., 04530 Mexico

[4] Department of Earth and Environmental Sciences, University of St Andrews, St Andrews, Scotland KY16 9AL

[5]NASA Ames Research Center, MS N245-3, Moffett Field, CA, 94035, USA

[6]Astronomy Department, University of Washington, Box 351580, Seattle, WA 98195-1580, USA

*Correspondence to: shawn.goldman@nasa.gov

**Abstract**: The search for life on planets outside our solar system will use spectroscopic identification of atmospheric biosignatures. The most robust remotely-detectable potential biosignature is considered to be the detection of oxygen ($O_2$) or ozone ($O_3$) simultaneous to methane ($CH_4$) at levels indicating fluxes from the planetary surface in excess of those that could be produced abiotically. Here, we use an altitude-dependent photochemical model with the enhanced lower boundary conditions necessary to carefully explore abiotic $O_2$ and $O_3$ production on lifeless planets with a wide variety of volcanic gas fluxes and stellar energy distributions. On some of these worlds, we predict limited $O_2$ and $O_3$ build up, caused by fast chemical production of these gases. This results in detectable abiotic $O_3$ and $CH_4$ features in the UV-visible, but no detectable abiotic $O_2$ features. Thus, simultaneous detection of $O_3$ and $CH_4$ by a UV-visible mission is not a strong biosignature without proper contextual information. Discrimination between biological and abiotic sources of $O_2$ and $O_3$ is possible through analysis of the stellar and atmospheric context – particularly redox state and O atom inventory – of the planet in question. Specifically, understanding the spectral characteristics of the star and obtaining a broad wavelength range for planetary spectra should allow more robust identification of false positives for life. This highlights the importance of wide spectral coverage for future exoplanet characterization missions. Specifically, discrimination between true- and false-positives may require spectral observations that extend into infrared wavelengths, and provide contextual information on the planet's atmospheric chemistry.

**Keywords:** astrobiology, biosignatures, exoplanets, low-mass stars, photochemistry, planetary atmospheres

## 1. Introduction

Various remotely-detectable biosignatures have been proposed for extrasolar planets, based on absorption features caused by atmospheric constituents that are predominantly or exclusively produced by life (Des Marais et al. 2002; Léger et al. 2011; Seager et al. 2012). Foremost among these gases are molecular oxygen ($O_2$) and ozone ($O_3$), particularly when either of these gases is present in an atmosphere that also contains methane ($CH_4$) (Des Marais et al. 2002; Lovelock 1965; Lederberg 1965). $O_2$ is a byproduct of oxygenic photosynthesis, the dominant primary producing metabolism on Earth; $O_3$ is produced photochemically in the atmosphere, and its concentration in Earth's atmosphere depends strongly on biological production of $O_2$. At least one of these two gases could potentially be detected by future missions such a proposed Occulting Ozone Observatory (O3, Pravdo et al. 2010; Savransky et al. 2010), and two probe-class Exoplanet mission currently under study, Exo-S (Starshade) and Exo-C (Coronograph) (Science and Technology Definition Teams, 2013). Both $O_2$ and $O_3$ would likely be detectable by the Advanced Technology Large Aperture Space Telescope (AT LAST) concept mission (Postman et al. 2008), by past TPF concept missions (Levine et al., 2006; Lawson et al., 2007), or potentially by the Exo-C (Stapelfeldt et al., 2014) and Exo-S (Seager et al., 2014) missions. Here, we deploy a broad sweep of relevant parameters to study the degree to which strictly abiotic processes might produce false positives for life by generating detectable amounts of $O_2$ or $O_3$. Based on the parameters for which this occurs, this allows a discussion of how the above missions might discriminate between any false positives and the true positives generated by life.

We do not anticipate direct abiotic sources of these gases, such as from volcanoes or from reactions at the sea floors of extrasolar planets. Thus, the primary abiotic source of both of these gases will be photochemistry, and will ultimately depend strongly on the availability of O atoms. The photochemical source of $O_2$ and $O_3$ is primarily controlled by the photolysis rates of $H_2O$, $CO_2$, $SO_2$ and other O-bearing gases. $H_2O$ can also deliver O atoms to the atmosphere, but very high $H_2O$ concentrations are controlled by climate and should present clear indications of the planet's lack of life, such as those produced by a runaway greenhouse planet (Schindler and Kasting, 2000). Furthermore, these features should be relatively short-lived, as this process leads to irreversible H escape to space, and sequestration of O in rocks at the surface. Once the planet's inventory of $H_2O$ is lost through this mechanism, the $O_2$ and $O_3$ concentrations will diminish. The photolysis rates of the other gases ($CO_2$, $SO_2$, etc.) are dictated by the planet's volcanic outgassing rates, and will be more sustainable on geological/astronomical timescales. $CO_2$ in particular has the potential to deliver a lot of O atoms to the atmosphere, as it is a major component of volcanic outgassing, is the dominant gas in the atmospheres of Mars and Venus, and has been proposed to be a major constituent of the past atmospheres of Earth and early Mars (Sagan and Mullen, 1972). O atoms can be liberated from $CO_2$ via photolysis:

$$CO_2 + h\nu\,(\lambda < 175\ \text{nm}) \rightarrow CO + O \qquad \text{(R1)}$$

Atomic O thus created through (R1) or photolysis of other O-bearing gases may recombine to form $O_2$, and eventually $O_3$. The distribution of those O atoms between $O_2$ and $O_3$ is critical to the concentration of either species, and is controlled by four reactions that are very well known from research on Earth's $O_3$ layer. This set of reaction is collectively known as the Chapman mechanism:

$$O_2 + h\nu\,(\lambda < 240\ \text{nm}) \rightarrow O + O \tag{R2}$$

$$O + O_2 + M \rightarrow O_3 + M \tag{R3}$$

$$O_3 + h\nu\,(\lambda < 340\ \text{nm}) \rightarrow O_2 + O \tag{R4}$$

$$O + O_3 \rightarrow O_2 + O_2 \tag{R5}$$

Here, '$h\nu$' represents photons of the indicated wavelength ($\nu = c/\lambda$, $c$ = speed of light), and the 'M' in reaction (R3) is a third molecule that only participates in the reaction to carry off excess energy, but is not consumed in the reaction. Because reactions (R1), (R2), and (R4) require photons of different energy levels (also, Figure 1), both the abundance and distribution of O atoms between O, $O_2$, and $O_3$ is subject to the wavelength-dependent stellar flux of the planetary host star.

$O_3$ concentrations should be particularly dependent on the wavelength distribution of the ultraviolet (UV) photons emitted by the host star (Figure 1). Far-UV (FUV, $\lambda < 200$ nm) photons drive $CO_2$ and $O_2$ photolysis and subsequent O production (R1), and therefore $O_3$ production (R2). By contrast, ozone destruction (R4) is primarily driven by mid-UV (MUV, 200 nm $< \lambda <$ 300 nm) photons, and can additionally be driven by near-UV (NUV, 300 nm $< \lambda <$ 440 nm) and visible (~440 to ~800 nm) photons (Sander et al. 2006). Because the sources and sinks of ozone drive the amount of $O_3$ in an atmosphere, both FUV ($O_3$ production) and MUV-NUV-visible radiation ($O_3$ destruction) will have a significant impact on $O_3$ concentrations. FUV photons are primarily produced by processes that correlate with stellar activity (Pace & Pasquini 2004), and MUV-NUV-visible photons are primarily generated from a star's blackbody radiation.

By definition, planets that are in the habitable zones of cooler stars absorb similar total amounts of energy as planets in the habitable zones of Sun-like stars (Kopparapu et al., 2013), but the wavelength distribution of that energy will be different. Cooler-type stars such as main sequence M stars (M dwarfs) produce relatively less NUV radiation than the Sun, but can produce comparable amounts of, or in some cases, more FUV radiation (Walkowicz et al. 2008; France et al. 2012; France et al. 2013). Hotter-type stars such as main sequence F stars have more radiation across the UV than the Sun, but this increase is more prevalent in the FUV. As a result, the FUV contributions to the stellar energy distributions of both M- and F-type stars can be much higher that that of the Sun, and planets in the habitable zones of these stars can accumulate greater amounts of atmospheric $O_2$ and $O_3$. This has been demonstrated for biologically mediated, oxygenated atmospheres similar to modern Earth (Selsis et al., 2002; Segura et al. 2003; Segura et al. 2005, Segura et al. 2010; Rugheimer et al., 2013). However, the most likely atmospheric composition for rocky habitable planets is $CO_2$, $H_2$ and $N_2$ (e.g. Zahnle et al., 2010; Seager and Deming, 2010). Selsis et al. (2002) were the first to study the potential for $O_2$ and $O_3$ to accumulate on planets devoid of life, but their work did not properly account for sinks of these gases (Segura et al., 2003). Tian et al. (2014) found a similar result using the spectrum of the M dwarf GJ876. Neither of those studies systematically studied the effects of atmospheric composition on the accumulation of detectable $O_2$ and $O_3$, and did not include hotter-type stars in the study. The lack of this parameter coverage limited the ability of these prior studies to discriminate between false and true positives for life.

Considering a wide range of possible planetary atmospheric compositions is critical, because sinks for $O_2$ and $O_3$ are primarily controlled by the chemical context of the atmosphere and oceans. In anoxic atmospheres, the greatest sinks for $O_2$ and $O_3$ are reactions with reduced radicals in the atmosphere, such as:

$$CH_3 + O_2 \rightarrow H_2CO + OH \quad \text{(R6)}$$

As the concentration of reduced species such as $CH_4$ increase in the atmosphere, so do the concentrations of radicals such as $CH_3$, and these should react with $O_2$ and $O_3$, keeping their concentrations low. It is therefore very difficult to maintain high levels of $O_2$, $O_3$, and $CH_4$ (or other reduced gases) in the atmosphere simultaneously. Major abiotic sources of reduced species include volcanic outgassing of $H_2$ and submarine production of $CH_4$, and their sinks are primarily determined by the redox state of the oceans. These are ultimately controlled by the redox state of the atmosphere and by the redox state of the oceans. Including the effects of the redox state of the oceans becomes critical for such simulations, and we developed a new methodology to ensure redox balance of the atmosphere-ocean system.

Abiotically-produced $O_3$ and $O_2$ concentrations are a sensitive function of $CO_2$ concentrations, stellar fluxes, and the fluxes of reduced species to the atmosphere. A photochemistry model that automates redox balance in the atmosphere-ocean system is needed to explicitly account for all these factors over a wide variety of parameter space, while producing accurate predictions of redox sensitive gas concentrations (Kasting, 2013). Below, we explain the development of such a model, and its application to the systematic exploration of abiotic productions of $O_2$ and $O_3$. We then simulate the spectral features that could arise from such atmospheres, and discuss the implications for exoplanet characterization mission concepts.

## 2. Atmospheric photochemistry model and redox balance automation

### 2.1 Photochemistry model

To investigate the effects of stellar energy distribution, bulk atmospheric composition, and boundary conditions on $O_3$ production for an abiotic terrestrial $CO_2$-$H_2O$-$N_2$ atmosphere, we used a photochemical model of prebiotic Earth. This code has been previously used to study abiotic Earth (Kharecha et al. 2005) as well as extrasolar terrestrial planets with biogenic gas fluxes around stars other than the Sun (Domagal-Goldman et al., 2011; Segura et al. 2003; Segura et al. 2005; Segura et al. 2007; Misra et al., 2014). Results from the photochemical code that exhibited potentially detectable biosignature gases were used as inputs to the Spectral Mapping Atmospheric Radiative Transfer (SMART) model (developed by D. Crisp), which produced simulated spectra for these model planets. SMART has been validated against observations of Earth, Mars, and Venus (Meadows & Crisp 1996, Crisp 1997; Halthore et al. 2005; Robinson et al. 2011; Robinson et al. 2014), and has been used to simulate spectra of extrasolar planets (Kiang et al. 2007; Domagal-Goldman et al. 2011; Robinson 2011).

The photochemical model computes the chemical equilibrium for 38 chemical species (Table 1) involved in 162 reactions, including but not limited to reactions (R1-R6). The atmosphere is divided into 100 layers with a fixed vertical extent of 1 km each. The long-lived chemical species are: O, $O_2$, $O_3$, $H_2O$, H, OH, $HO_2$, $H_2O_2$, $H_2$, CO, $CO_2$, HCO, $H_2CO$, $CH_4$, $CH_3$, $C_2H_6$, NO, $NO_2$, HNO, SO, $SO_2$, and $H_2SO_4$; $N_2$ is also included with a constant mixing ratio. The input stellar flux for the model includes the interval from 130 nm to 855 nm, and also includes the Lyman α flux at 121 nm. Detailed model descriptions can be found in Kharecha et al. (2005); Segura et al. (2007) and Guzmán-Marmomejo et al. (2013).

**Table 1**
Species in photochemical model, with default boundary conditions and redox contributions.

| Species | Lower bound type | Lower bound value | Reducing power ($H_2$) |
|---|---|---|---|
| O | $v_{dep}$ | 1. | -1 |
| $O_2$ | $v_{dep}$ | 0-1.5×$10^{-4}$ | -2 |
| $H_2O$ | - | - | 0 |
| H | $v_{dep}$ | 1. | 0.5 |
| OH | $v_{dep}$ | 1. | -0.5 |
| $HO_2$ | $v_{dep}$ | 1. | -1.5 |
| $H_2O_2$ | $v_{dep}$ | 0-0.2 | -1 |
| $O_3$ | $v_{dep}$ | 0-1×$10^{-3}$ | -3 |
| $H_2$ | $v_{dep}$ | 0-2.4×$10^{-4}$ | 1 |
| $CO_2$ | $fCO_2$ | 0.05 | 0 |
| CO | $v_{dep}$ | 0-1.2×$10^{-4}$ | 1 |
| HCO | $v_{dep}$ | 1. | 1.5 |
| $H_2CO$ | $v_{dep}$ | 0-0.2 | 2 |
| $CH_4$ | *flux* | 0-6.8×$10^8$ | 4 |
| $CH_3$ | $v_{dep}$ | 1. | 3.5 |
| $C_2H_6$ | $v_{dep}$ | 0. | 7 |
| NO | $v_{dep}$ | 0-3. ×$10^{-4}$ | -1 |
| $NO_2$ | $v_{dep}$ | 0-3. ×$10^{-3}$ | -2 |
| HNO | $v_{dep}$ | 1. | -0.5 |
| SO | $v_{dep}$ | 0-3. ×$10^{-4}$ | 1 |
| $SO_2$ | $v_{dep}$ | 1. | 0 |
| $H_2SO_4$ | $v_{dep}$ | 1. | -1 |

Default starting boundary conditions are either fixed deposition effiency ("$v_{dep}$"), constant mixing ratio ("$fCO_2$"), or constant flux ("*flux*"); the first two quantities are dimensionless, *fluxes* are in molecules/cm$^2$/s. Ranges are for species whose boundary conditions were allowed to change to ensure redox balance in the oceans, as explained in section 2.2. Upper limits for these species were determined by the maximum rate they could diffuse into an empty ocean, the minimum rate was 0., representing an ocean saturated with that species. Reducing power is measured in units scaled to the reducing power of an $H_2$ molecule (Kharecha et al. 2005).

No photosynthetic source of oxygen is included in our photochemical model; the boundary conditions instead reflect a planet devoid of life (Table 1, Section 2.2). Only H, CO, O, and $CO_2$ were allowed to cross the top layer of the atmosphere; all other species had a "0 flux" boundary condition at the top of the uppermost layer of the atmosphere. H was allowed to flow up and out through the top of the atmosphere, in accordance with diffusion-limited escape of H to space. $CO_2$ flow past the top layer of the model was converted to an equal influx of CO and O, consistent with $CO_2$ photolysis above the top of our model grid. At the lower boundary, $CH_4$ was assigned a surface flux from abiotic production in the oceans resulting from serpentinization; and $H_2$ was given both a volcanic outgassing rate and a fixed deposition velocity into the oceans. The $CO_2$ surface mixing ratio and $H_2$ outgassing rate varied between different simulations to cover a variety of planetary conditions (Figure 2). For all other species, we assumed reaction with the surface or dissolution into oceans at rates proportional to the solubility of each species (Kharecha et al. 2005). For soluble species, we allowed the dissolution rates to go down from their dissolution limited rates in order to maintain ocean redox balance, as explained in the next section.

**2.2 Redox Balance Automation**

Redox balance between oxidants and reductants in the surficial environment should be the primary driver of the atmospheric chemistry of a potentially habitable but lifeless world. Any redox imbalance in the surface environment of a planet will lead to changes in reaction rates or removal of species that would drive the system back towards a dynamic, redox-balanced equilibrium. For example, if a planetary surface had an excess of oxidants, it would lead to greater rates of reactions that destroy these oxidants, or burial of minerals that would remove these oxidants from the surface. Through either pathway, their concentrations would be decreased, thereby driving the system back towards redox balance. As a result, any such imbalance would be short-lived and less likely to be observed. Conversely, a redox-balanced simulation represents a point of "dynamic equilibrium" that would be the most likely state for a planet with the given outgassing rates, bulk composition, and physical properties. Note this dynamic equilibrium is not the same thing as chemical equilibrium due to the energy input from stars, but instead represents a steady-state atmospheric composition.

The one thing that can change the redox balance of an Earth-mass planet is hydrogen escape to space, which can irreversibly change the net redox inventory of the surface and near-surface environment (see Catling et al. 2014 for a review of this concept, and on redox controls in Earth's history). (For smaller planets, escape of other species, such as C and O, could also affect redox balance.) However, on timescales less than the hundreds of millions of years required for the escape process to affect the bulk redox of the surface and upper mantle, the surface and atmospheric redox state will be buffered to match the redox state of the mantle. In other words, escape can slowly change the redox state of the planet, but at any given point in the planet's evolution, the redox states of the atmosphere, the ocean, and the atmosphere-ocean system should be balanced. The simulations presented here can be considered representations of a planet in such a steady state, with models that have smaller contributions of reduced species from the subsurface simulating planets that are further along in their oxidation via atmospheric H escape.

Redox balance in this model can be tracked in units of "$H_2$ reducing power," measured by the amount of $H_2$ required to convert the redox-neutral gases (defined in our calculations as $H_2O$, $CO_2$, $N_2$, and $SO_2$ for H and O, C, N, and S, respectively) into the species in question. For example, atmospheric $CH_4$ can be produced by the representative reaction $CO_2 + 4\ H_2 \rightarrow CH_4 + 2H_2O$; therefore, the addition of one $CH_4$ molecule to an atmosphere is equivalent to the addition of four molecules of $H_2$. We can use this scheme to tabulate the sources and sinks of $H_2$ for each species in the atmosphere (Table 1), in the oceans, and across the ocean-atmosphere boundary.

In past work (*e.g.*, Segura et al., 2007), the focus was on the redox balance in the atmosphere, between the volcanic outgassing of reduced species, escape of hydrogen (a reduced species) from the top of the atmosphere, the flux of methane (a reduced species) from the ocean into the atmosphere, and any redox imbalance in the deposition of gases to the surface and oceans. This can be expressed in an equation showing the balance of sources and sinks of reductants:

$$F_{volc} + F_{CH_4} + F_{ox\ dep} = F_{esc} + F_{red\ dep}, \text{ (E1)}$$

where $F_{volc}$ is the flux of reduced volcanic species, $F_{CH_4}$ is the flux of methane from the oceans, $F_{ox\ dep}$ is the deposition flux of oxidized species (removal of oxidants is equivalent to addition of reductants), $F_{esc}$ is the rate of escape of reductants out the top of the atmosphere, and $F_{red\ dep}$ is the rate of deposition of reductants. All of these are tracked in units of "$H_2$ reducing power" as explained in the preceding paragraph.

However, a more careful redox balance also must include reactions that occur in the oceans into which the atmospheric species are being deposited. We assume reactions with land surfaces get propagated to the oceans through rivers, so all species deposited at the surface end up in this ocean budget. With this assumption, the redox balance is between the flux of reduced species from the atmosphere, the flux of oxidized species from the atmosphere, the flux of methane to the atmosphere, and any redox imbalance at the ocean floor, including deposition of oxidized species such as iron-oxides, deposition of reduced species such as organic carbon, and the flux of reduced species such as $Fe^{2+}$ into ocean water from the subsurface. The net reaction for ocean redox balance is therefore (again measured in units of "$H_2$ reducing power"):

$$F_{red\ dep} + F_{floor} = F_{ox\ dep} + F_{CH4}, \text{ (E2)}$$

where $F_{floor}$ is the redox imbalance at the ocean floor, and $F_{red\ dep}$, $F_{ox\ dep}$, and $F_{CH4}$ represent the same values as they do in equation (E1).

$F_{floor}$ is effectively a free model parameter, for which a value must be chosen. We used values from $3\times10^{9}$-$3\times10^{11}$ molecules cm$^{-2}$ s$^{-1}$, using estimates for the iron oxide deposition rate on early Earth ($5\times10^{11}$ molecules cm$^{-2}$, from Holland, 1984) as an upper limit, with lower values simulating more oxidized planets with a higher propensity for accumulation of abiotic $O_2$ and $O_3$. These lower values of $F_{floor}$ may be appropriate for planets with an oxidized bulk composition (compared to Earth), or for water-rich planets with high enough seafloor pressures that an ice VII layer forms at the ocean floor, potentially decreasing the flux of reduced mantle materials (Léger et al., 2004).

The photochemical model automatically ensures equation (E1) is satisfied, for a given set of boundary conditions (Kharacha et al., 2005). To ensure equation (E2) was additionally satisfied for all our simulations, we created a script that changes the model's boundary conditions between simulations, and repeatedly runs the photochemical code until equation (E2) was satisfied with the given value of $F_{floor}$.

To achieve this, we used a "piston velocity" treatment of the diffusion of gases into the ocean (Kharecha et al., 2005). This treatment traces its origins to the biogeochemical work of Broecker and Peng (1982) that describes and explains the behavior of elements in Earth's oceans. Under this treatment, it is assumed that the rate at which gases diffuse into the ocean can be approximated with a conceptual model of a piston pushing gases through a thin layer into the ocean. The velocity of that piston for a given gas, X, is given by $v_p(X) = K_{diff}(X)/z_{film}$, where $v_p(X)$ is the piston velocity of gas X, $K_{diff}(X)$ is the thermal diffusivity of gas X, and $z_{film}$ is an empirically-determined surface layer thickness (40 μm) that is the same for all gases. This approach is used to approximate multiple, complex phenomena, including wave breaking, bubble formation, and molecular diffusion. It yields an equation for the rate at which gas X dissolves into the ocean:

$$\phi(\mathrm{X}) = v_p(\mathrm{X})\cdot(\alpha(\mathrm{X})\cdot p\mathrm{X} - [\mathrm{X}]_{aq})\cdot C, \qquad (E3)$$

where $\phi(\mathrm{X})$ is the rate at which gas X flows into the ocean, $v_p(\mathrm{X})$ is the piston velocity for gas X, $\alpha(\mathrm{X})$ is the solubility of gas X in water, $p\mathrm{X}$ is the partial pressure of gas X at the bottom of the atmosphere, $[\mathrm{X}]_{aq}$ is the concentration of gas X at the top of the ocean, and $C$ is a conversion factor ($6.02\cdot10^{20}$ molecules $\mathrm{cm}^{-3}\cdot\mathrm{mol}^{-1}\cdot\mathrm{L}$).

At the start of each set of runs, we assumed the ocean was empty of all soluble species ($[\mathrm{X}]_{aq} = 0$), which gives the maximum rate at which any species can diffuse into the oceans:

$$\phi_{max}(\mathrm{X}) = v_p(\mathrm{X})\cdot\alpha(\mathrm{X})\cdot p\mathrm{X}\cdot C. \qquad (E4)$$

If equation (E2) was not satisfied, we ran the photochemical model again, assuming the ocean would start to “fill up” with the soluble species contributing the most to the model ocean’s redox imbalance: if the ocean was too oxidized, we lowered the deposition velocity of the gas contributing the most to the ocean’s oxidizing power; conversely, if the ocean was too reduced, we lowered the deposition velocity of the gas contributing the most to the ocean’s reducing power. We set a minimum diffusion velocity of 0 for all soluble species; once a species reached this minimum deposition rate the next-greatest contributor to the redox balance was altered. This approach is consistent with $[\mathrm{X}]_{aq}$ increasing from its minimum value of $[\mathrm{X}]_{aq} = 0$, up to its maximum value of $[\mathrm{X}]_{aq} = \alpha(\mathrm{X})\cdot p\mathrm{X}$. Higher concentrations of X require production of X in the oceans, and (except CH4, which is discussed below) this case is reserved for systems with biological production of X.

Highly reactive species (H, OH, O, HCO, $CH_3$, HNO) were excluded from this process, as these species would rapidly react and never become saturated in the ocean, so we assumed the ocean is always empty of these species. Two additional species ($SO_2$ and $H_2SO_4$) are highly soluble, so we assumed the ocean would not fill with them, either. That left 9 species whose deposition velocities were allowed to change between model iterations: $O_2$, $H_2O_2$, $O_3$, $H_2$, CO, $H_2CO$, NO, $NO_2$, and SO. If changes to the deposition rates of these species affected the net redox state of the ocean by less than 1%, we instead altered the $CH_4$ flux from the ocean to the atmosphere, within bounds of 0 and $6.8\times10^8$ molecules/cm$^2$/s, the estimate for the Earth's global prebiotic $CH_4$ flux from serpentinization (Guzmán-Marmolejo et al., 2013). We also ran simulations where all 17 species were allowed to vary as if the ocean could fill up with them. Doing this affected the quantitative predictions of $O_2$ and $O_3$ concentrations, but the effect of these changes were not large enough to qualitatively impact the absorption features that could be remotely observable.

Although it is theoretically possible for higher $CH_4$ fluxes to originate from a planet with a different composition, accurately calculating them requires a separate geochemical model that currently does not exist. We repeated the process of running the model and iterating on the boundary conditions until the net ocean-atmosphere system obtained redox balance. The criterion we used for this "global" redox balance was that there was not a net reduced flux to the oceans, and that any net oxidant flux to the oceans was smaller than that simulation's proscribed limit on $F_{floor}$. This is equivalent to assuming the mantle provides the ocean with some non-zero, finite flux of reductants that can serve as a sink for oxidants from the atmosphere.

## 3. Parameter Ranges of Simulated Atmospheres

As described above (SECTION 1), the major controls on the amount of photochemically-produced $O_2$ and $O_3$ in an abiotic atmosphere are: 1) the bulk atmospheric composition; 2) the fate of atmospheric species at the interface with continents and oceans; and 3) the quantity and wavelength-distribution of UV radiation from the host star. We varied all of these factors by running our model for different stellar spectra, $CO_2$ mixing ratios, $H_2$ outgassing rates, and different values of $F_{floor}$. (This last term can be thought of as different rates of $Fe^{2+}$ upwelling or, equivalently, different rates of Fe-oxide deposition.) The permutations of these boundary conditions led to a total of 14,499 converged simulations of abiotic atmospheres that maintained redox balance in the ocean-atmosphere system.

### 3.1 Parameter Ranges of Photochemistry Simulations

To study the effects of the bulk atmospheric composition, we varied the $CO_2$ surface mixing ratio from $3\times10^{-4}$ (300 ppm, close to the modern Earth's value) to $5\times10^{-1}$ (50% of the atmosphere) and varied the $H_2$ outgassing rate from $2\times10^{6}$ to $2\times10^{12}$ molecules/cm$^2$/s, bracketing the modern value of $2\times10^{9}$ molecules/cm$^2$/s by three orders of magnitude. We parameterized $F_{floor}$, the flux rate of reduced species (such as $Fe^{2+}$) into the oceans, over two orders of magnitude, from $3\times10^{9}$ to $3\times10^{11}$ molecules $Fe^{2+}$/cm$^2$/s, with the top end of that range similar to the estimated rate of Fe-oxide deposition on Earth prior to the rise of oxygen ($5\times10^{11}$ molecules Fe/cm$^2$/s).

### 3.2 Properties of Stars Used in Simulations

To study the effects of the stellar energy distribution, we simulated planets with an incoming solar spectrum that would be consistent with a planet in the habitable zones of Sigma Boötis (an F2V-type star), the Sun, ε Eridani (a K2V-type star), AD Leonis (M3.5V-type star) and GJ876 (M4V-type star). The characteristics of stars used for the present simulations are presented in Table 2. The F and K stars have been used by several authors to study biosignatures in atmospheres similar to present Earth, with high $O_2$ and low $CO_2$ (*e.g.* Segura et al. 2005; Grenfell et al 2007; Rugheimer et al. 2013). For the M dwarfs we used spectra from AD Leonis (AD Leo, Table 2) and GJ 876. AD Leonis is a star that has been extensively used in studies of planetary atmospheres (e.g. Domagal-Goldman et al. 2010; Segura et al. 2010; Rauer et al. 2011; von Paris et al. 2013), and GJ 876 is one of the M stars that have UV flux measurements (France al. 2013). The AD Leonis spectrum is the same as the one used in Segura et al. (2005, 2010). To compile the GJ876 stellar input spectrum,  we use the GJ 876 spectrum reported by France et al. (2012) from 130 to 320 nm combined with a NextGen model (v5) star with $T_{eff}$ = 3200 K (http://hobbes.hs.uni-hamburg.de/~yeti/NG-spec.html) from 320 to 855 nm.

Reviews on the potential habitability of planets around M dwarfs (e.g. Scalo et al. 2007, Tarter et al. 2007) made clear that many properties of these stars are not fully known and thus their effects on planetary habitability have not been constrained. Particularly, chromospheric activity and its resultant UV emission are not fully understood and pose a complex problem for stellar atmospheric models (e.g. Walkowicz et al. 2008), and Segura et al. (2005) demonstrated that the slope of the UV emission from M dwarfs was relevant to the atmospheric chemistry of biosignatures as it affects destruction rates and therefore atmospheric lifetimes of potential biosignatures gases. M dwarfs are classified as active based on their H$\alpha$ strong chromospheric emission while those with H$\alpha$ in absorption were "quiescent", although their UV spectra were unknown and therefore it was not possible to assess if their chomospheres were inactive. UV spectra from very active red dwarfs are characteristically constant in the wavelength range from 100 to 300 nm. Observations from the *Hubble Space Telescope* (Walkowicz et al. 2008; France et al. 2012; France et al. 2013) showed that the UV emission from stars classified as inactive based on their Hα emission was similar to that of the active M dwarfs; therefore, the characteristic flat UV spectra of active M dwarfs when quiescent (no flaring) may be representative of non-active M dwarfs as well. In particular, GJ 876 shows Hα in absorption but it presents a measurable emission in X-rays, a characteristic of chromospherically active stars (France et al. 2012 and references therein). Observations with the Hubble Space Telescope showed that the near UV emission (200 to 320 nm) of GJ 876 was actually similar to that of AD Leonis (Walkowicz et al. 2008). For the relatively blue stars (F dwarfs), their high amounts of UV radiation are caused by blackbody radiation extending into the MUV, as opposed to chomospheric activity at FUV wavelengths (Figure 1).

For all stars, we placed the model planets in an orbit that would allow the surfaces of these planets to absorb approximately the same amount of energy as the Earth absorbs from the sun. This includes a correction factor for the broadband albedo of a planet caused by the interaction of a planet's albedo spectrum with the stellar energy distribution. This was done after the process adopted by Segura et al. (2005).

**Table 2.**
Stellar parameters

| Star | Spectral type[a] | Star $T_{effective}$ (K)[b] | Luminosity ($L_{\odot}$)[c] | Star distance (pc)[d] | Semimajor axis (AU)[e] | Lyman-α flux (erg $cm^{-2}$ $s^{-1}$)[f] |
|---|---|---|---|---|---|---|
| σ Boötis (HD128167) | F3V*wv ar* | 6770 | 3.18 | 15.6 | 1.69 | 29.64 |
| Sun | G2V | 5770 | 1 | - | 1 | 8.17 |
| ε Eridani (HD 22049) | K2V | 5000 | 0.33 | 3.2 | 0.59 | 159.13 |
| AD Leonis (GJ 388) | M3.5e | 3400 | 0.023 | 4.9 | 0.16 | 390 |
| GJ 876 (HIP 113020) | M4V | 3172 | 0.013 | 4.6 | 0.12 | 28.5 |

[a] K and F stars: Habing et al. (2001). AD Leonis: Hawley and Pettersen (1991). GJ 876: Reid et al. (1995).
[b] K and F stars: Habing et al. (2001). AD Leonis: Leggett et al. (1996). GJ 876: Jenkins et al. (2009)
[c] K and F stars: Valenti and Fischer (2005). AD Leonis: Leggett et al. (1996). GJ 876: Delfosse et al. (1998)
[d] K and F stars: Habing et al. (2001). AD Leo: Cruz and Reid (2002). GJ 876: Reid et al. (1995).
[e] Calculated as described in Segura et al. (2003, 2005)

[f] Flux received at the planet. F and K stars: Landsman and Simon (1993). Sun: World Metereological Organization (1985). AD Leonis: Segura et al. (2005). GJ 876: France et al. (2012)

In this study, we limit ourselves to stellar spectra that have been observed, or modeled based on other observations. While it is possible that extremely active stars could contribute significant energy to $CO_2$ photolysis and $O_2$/$O_3$ production (K. Zahnle, personal communication), such stars would also deliver significant energy for $O_2$/$O_3$ destruction in the form of energetic protons. Past modeling of both these effects show that, if anything, the net impact of flares would be to diminish, not enhance, $O_3$ concentrations (Segura et al., 2010), and should only be important for very active stars. Another possibility for enhanced $O_2$ and $O_3$ production beyond the parameter space explored here would be planets around O, B, and A stars that have significant UV radiation even when the stars are quiescent. However, due to the relatively short lifetimes of habitable zones around these objects, they are not typically included in target lists for future exoplanet characterization missions (Turnbull et al., 2003).

## 4. Results and Discussion

Figure 2 shows color contours of $O_2$ and $O_3$ column densities as a function of $CO_2$ mixing ratio at the surface, $H_2$ outgassing rates, seafloor fluxes of reduced material to the oceans, and star type. Both $O_2$ and $O_3$ column densities are controlled by the atmospheric and stellar context – for either species, the highest values occur for high $CO_2$ concentrations, low $H_2$ outgassing rates, low flux rates of reduced materials into the oceans, and high stellar FUV fluxes. This makes sense in the framework of source/sink rationale. As $CO_2$ concentrations or FUV fluxes increase, reaction R1 goes faster, speeding up the production of O atoms for $O_2$ formation. As $H_2$ outgassing rates decrease, so does the sink for O atoms from reactions with reduced radicals, for example through reaction (R6). The effect of decreasing the flux rates of reduced material into the oceans has a similar effect, albeit mediated by the oceans and ocean-atmosphere interface.

We will spend the rest of this manuscript studying the atmospheres with the highest abiotic $O_2$ and $O_3$ concentrations in more detail. Table 3 shows the total inventories of $O_2$ and $O_3$ for the six cases we will focus on. These include the pre-industrial Earth as a control to compare against (Metz et al. 2007), as well as photochemical simulation of abiotic planets in the habitable zones of σ Boötis (F2V), ε Eridani (K2V), AD Leonis (M3.5V), and GJ 876 (M4V). These photochemical models (Table 3) have 5% $CO_2$, a volcanic $H_2$ flux of $1\times10^6$ molecules·cm$^{-2}$·s$^{-1}$, and an ocean iron flux of $1\times10^{10}$ molecules·cm$^{-2}$·s$^{-1}$. The $O_3$ column depth for the abiotic model planets around the M dwarf AD Leonis and the F dwarf σ Boötis (respectively $9\times10^{15}$ and $1\times10^{18}$ molecules cm$^{-2}$) were significantly greater than the value for the model of abiotic Earth around the Sun ($4\times10^{15}$ molecules cm$^{-2}$), but significantly less than Earth's modern $O_3$ column depth ($9\times10^{18}$ molecules cm$^{-2}$). These $O_3$ column depths are also greater than the values (5-$10\times10^{14}$ molecules·cm$^{-2}$) recently detected on Venus by the Venus Express mission (Montmessin et al. 2011). For $O_2$, the column depth is an order of magnitude higher on the model planet around the F-type star σ Boötis ($1\times10^{21}$ molecules cm$^{-2}$) than any of the other model planets, but orders of magnitude lower than the $O_2$ column on pre-industrial Earth ($5\times10^{24}$ molecules cm$^{-2}$), and lower than the largest values reported by Tian et al. (2014), who report column densities up to $1\times10^{23}$ molecules cm$^{-2}$.

**Table 3**

Ozone and oxygen column depths for simulated atmospheres

| Planet | Star | $CO_2$ mixing ratio (ppm) | $CH_4$ mixing ratio (ppm) | $O_3$ column depth (cm$^{-2}$) | $O_2$ column depth (cm$^{-2}$) |
|---|---|---|---|---|---|
| Present Earth[a] | Sun (G2V) | 335 | 1.79 | $8.6\times10^{18}$ | $4.6\times10^{24}$ |
| Abiotic Earth[b] | σ Boötis (F2V) | 500,000 | $3.6\times10^{-4}$ | $1.3\times10^{18}$ | $1.7\times10^{21}$ |
| Abiotic Earth[b] | Sun (G2V) | 500,000 | $2.0\times10^{-3}$ | $4.5\times10^{15}$ | $9.9\times10^{19}$ |
| Abiotic Earth[b] | ε Eridani (K2V) | 500,000 | $1.0\times10^{-2}$ | $1.0\times10^{15}$ | $3.4\times10^{19}$ |
| Abiotic Earth[b] | AD Leonis(M3.5V) | 500,000 | $9.4\times10^{-3}$ | $8.8\times10^{15}$ | $7.4\times10^{19}$ |
| Abiotic Earth[b] | GJ 876 (M4V) | 500,000 | $8.5\times10^{-2}$ | $1.8\times10^{15}$ | $2.7\times10^{19}$ |

[a]pre-industrial Earth (Metz et al. 2007)

[b]photochemical simulations explained in manuscript.

Whether or not these potential "false positives" for life present an issue for future planet characterization missions depends on our ability to detect features from abiotically-generated $O_2$ and $O_3$, and upon our ability to discriminate such false positives from those generated by an inhabited planet. We discuss detectability first, and then requirements for false positive discrimination.

**4.1 Comparisons to Prior Work**

At least three other groups (Selsis et al., 2002; Hu et al., 2012; Tian et al., 2014) have previously reported accumulation of detectable $O_2$ and $O_3$ in planetary atmospheres using altitude-dependent photochemical models. However, the first of these studies (Selsis et al., 2002) did not balance redox in the atmosphere, which can lead to spurious accumulation of atmospheric $O_2$ and $O_3$ by neglecting sinks for these species via reaction with reductants at the planet's rocky surface or in its oceans (Segura et al., 2007).

The other two studies (Hu et al., 2012; Tian et al., 2014) produced high $O_2$ and $O_3$ while maintaining redox balance, but did so with boundary conditions that we contend are not as rigorous as those employed here. Specifically, the deposition velocities of $H_2$, CO, and $O_2$ adopted by these two groups may have enhanced $O_2$ and $O_3$ concentrations, either by impacting the bulk redox budget, or by impacting the distribution of oxidized and reduced species in the atmosphere. This allows them to satisfy equations E1 and E2 above, but in a biased fashion that can favor accumulation of certain oxidized species over others. (For that matter, it also can favor accumulation of certain reduced species over others). Hu et al. (2012) proscribe $H_2$, CO, and $O_2$, deposition velocities to be 0, $10^{-8}$, and 0 cm/s, respectively; they cite Kharecha et al. (2005) for the value for CO, who base the CO deposition velocity on the chemistry of CO in water, and assume the oceans are saturated with respect to $H_2$ and $O_2$. This effectively eliminates one of the most important sinks for $O_2$ in an abiotic world: reactions with surface rocks and dissolved species in ocean water.

In Tian et al. (2014), the deposition velocities proscribed for $H_2$, CO, and $O_2$ were respectively 0, $10^{-6}$, and $10^{-6}$ cm/s. The rationale for this is that $H_2$ would not be consumed by biology in a prebiotic ocean (and therefore the oceans would be saturated in $H_2$), and that CO and $O_2$ would be consumed by redox reactions in the oceans. Both this approach and that of Hu et al. (2012) are based on assumed CO-consuming reactions in the ocean, but the soluble species that destroy CO should also act to destroy other reduced species. Not accounting for this could overestimate the rate at which CO flows into the oceans, underestimate atmospheric CO concentrations, and underestimate the rate at atmospheric CO reacts with $O_2$ and $O_3$. A more comprehensive and less selective approach is to manage the deposition velocities of $H_2$, CO, $O_2$ – and all other soluble species – based on lowering ocean concentrations of the species contributing the most to ocean redox imbalance. This can be done in a manner that is consistent with the fundamental assumptions in the deposition velocity model, as described above (section 2.2).

When the different boundary conditions are taken into account, we are able to replicate prior work, with the exception of the Tian et al. (2014) results. When we adopt the boundary conditions of Hu et al. (2012), we replicate their results. We also replicate those in Segura et al. (2007) for cases with similar stellar fluxes, $H_2$ outgassing rates and surface $CO_2$ mixing ratios. However, we were not able to replicate the results of Tian et al., (2014) using photochemical models (Domagal-Goldman et al., 2010; Guzmán-Marmomejo et al., 2013; Kurzweil et al., 2013) that share a common heritage and a similar chemistry network with their model, even when using the boundary conditions given in Tian et al. (2014) and the same source for the stellar flux (France et al., 2012). In general, using similar methodologies we predict at least an order of magnitude lower $O_2$ column densities than Tian et al. (2014) for all our models. If we adopt our boundary condition method (section 2.2), we find orders of magnitude lower $O_2$ concentrations. For example, for a 5% $CO_2$ atmosphere with the same $H_2$ outgassing rates as Tian et al. (2014), we predict an $O_2$ column depth of $8\times10^{18}$ molecules/cm$^2$, compared to the Tian et al. (2014) value of $8\times10^{22}$ molecules/cm$^2$.

**4.2 Detectability of Potential Biosignature Features in Atmospheres of Abiotic Planets**

We used our radiative transfer model (SMART) to simulate the reflection and emission spectra of the six planets shown in Tables 3 and 4. Figures 3−5 show spectra in the UV, visible−near-infrared, and mid-infrared, respectively. For comparison, the top panels of each of these figures shows a spectrum of modern Earth, generated by the well-validated Virtual Planetary Laboratory 3-D spectral Earth model (Robinson et al. 2010; Robinson et al. 2011; Robinson et al. 2014).

For all the model planets, the simulated reflectance spectra show the presence of the $O_3$ Hartley bands, which are seen in absorption from 0.2-0.3 μm (Figure 3). $O_3$ absorption is detectable at the 3-sigma level for abiotic planets around σ Boötis (F2V), the Sun (G2V), ε Eridani (K2V), AD Leonis (M3.5V), and GJ 876 (M4V) for missions that can respectively achieve a signal-to-noise of 3, 16, 64, 8, and 38 in the UV (Table 4). However, these $O_3$ signals for many of these model planets will be much weaker than the $O_3$ feature caused by a biosphere. It will therefore be distinguishable if the bottom of the absorption feature in the albedo spectrum can be measured. A stronger $O_3$ absorption feature is present in the simulated albedo spectrum of the model planet around σ Boötis (an F2V star), making the detection of the presence of $O_3$ relatively easy for any mission that can detect radiation in the UV wavelength range, such as the Occulting Ozone Observatory (O3). The shape of the feature is also very similar to the one in Earth's modern spectrum, meaning that a relatively high spectral resolution is required to differentiate between the $O_3$ levels for the abiotic model planet and inhabited Earth. Our simulation of the abiotic planets around sigma Boötis also exhibits an $O_3$ feature at 9.6um (Figure 5), which overlaps strong absorption by the doubly hot band of $CO_2$ at 9.4um. The ozone absorption is more cleanly seen when the absorption from $CO_2$ is removed from the spectrum (the grey spectrum in panel 2 of Figure 5), as could be done by modeling and removal during analysis of an observed spectrum. Detecting this feature, once the carbon dioxide has been modeled and removed, requires a SNR > 15 (Table 4). The abiotically generated ozone predicted for the σ Boötis planet would not likely be detectable by JWST. However, if this instrument is used to characterize transit spectra of any potentially-habitable super-Earth planets (Kaltenegger and Traub, 2009), the cautions above should be taken into consideration.

**Table 4.**

Calculated Signal-to-Noise Required (SNR) to detect absorption features of abiotic planets at a 3σ confidence level, for a mission with a spectral resolution of $\lambda/\Delta\lambda$ =75. This is calculated from SNR = $3\times F_{featureless}/(F_{featureless}-F_{complete})$, at the wavelength in question, where $F_{featureless}$ is the spectrum without the contributions from the gas causing the feature, and $F_{complete}$ is the full spectrum, including contributions from that species. The factor of 3 is to ensure a 3σ confidence level.

| Star | 0.25 μm $O_3$ | 0.76 μm $O_2$ | 1.7 μm $CH_4$ | 2.3 μm CO | 3.3 μm $CH_4$ | 4.6 μm CO | 9.6 μm $O_3$ |
|---|---|---|---|---|---|---|---|
| σ Boötis (F2V) | 3 | 380 | 570 | 8.6 | 1300 | 3.6 | 15 |
| Sun (G2V) | 16 | 2800 | 430 | 28 | 580 | 11 | 3000 |
| ε Eridani (K2V) | 64 | 1600 | 95 | 44 | 120 | 21 | 2100 |
| AD Leo (M3.5V) | 7.5 | 620 | 100 | 27 | 120 | 13 | 1000 |
| GJ 876 (M4V) | 38 | 6600 | 17 | 33 | 25 | 24 | 2700 |

$CH_4$ features are potentially detectable on the model planet around GJ 876, given measurements with a SNR above 17 at 1.7 μm and complete removal of the overlapping $CO_2$ absorption feature that dominates at this wavelength, or a SNR above 25 at 3.3 μm, with independent limits on $CO_2$ and CO concentrations. However, no $O_3$ or $O_2$ feature can be detected for the same planet for a SNR <30. The reason for the higher concentrations of $CH_4$ on GJ 876 (compared to models of planets around other stars) is the lower $O_2$ concentrations, and a correspondingly lower sink for $CH_4$. The first two of these features could be detectable by a flagship-class mission with a UV-NIR wavelength range that can also observe the $O_3$ feature at 0.25 μm. This combination could constitute a false positive for life, without a strategy (which could include NIR observations) to discriminate its abiotic origin.

Absorption features from other biosignature gases in these atmospheres would be much more difficult to detect (Table 4). The next strongest $O_2$ feature occurs at 0.76 μm, and for the relatively low abiotically produced $O_2$ abundances, this gas is only detectable with measurements that have a SNR above 380. This is an extremely difficult measurement to make, and therefore is likely not a concern for proposed exoplanet characterization missions. We did not include any $O_2$ dimer features, as the spectral model used here pre-dates the improvement needed to study them (Misra et al., 2014).

**4.3 Discrimination of Living Planets from Abiotic Planets with $O_3$ Absorption Features**

By observing the atmospheric context more completely, discrimination between biological and abiological sources for the $O_3$ is possible. Ideally, this would involve observations of the stellar spectrum, broad-wavelength planetary spectra sufficient to constrain other gas concentrations, and detailed atmospheric models similar to the ones used in this study. Knowledge of the UV stellar energy distribution is needed to provide inputs to photochemical models that could calculate the fluxes needed to maintain the concentrations of various gases derived from the spectrum. These models could then attempt to constrain a number of spectral features that are particular to an atmosphere with abiotic $O_3$.

First, the quantification of the abundance of $O_3$ could indicate atmospheric (and abiotic) $O_3$ production, as the $O_3$ column density in all our simulations of abiotic planets is at least an order of magnitude less than Earth's modern $O_3$ column density (Table 3). For most of the abiotic $O_3$ concentrations modeled here, quantification would require measurement of the depth of the 0.25 μm $O_3$ absorption feature, which is not saturated, and exhibits at least a small amount of transmission through the planet's atmosphere. The lone exception is the abiotic planet simulated around σ Boötis, which, like modern Earth, exhibits a spectrum that is completely opaque from 0.23-0.31um. This case would be much more difficult to discriminate as abiotically-generated, as the saturated ozone band would give only a lower limit on the ozone abundance in the atmosphere. It would not be possible to tell if the actual abundance was much higher, and therefore potentially biologically generated.

Second, a mission could discriminate between biotic and abiotic sources of $O_3$ by measuring the $O_3/O_2$ ratio of the atmosphere. Despite detectable $O_3$ concentrations, in all cases there was a lack of an observable $O_2$ absorption feature at 0.76 μm. The lack of this $O_2$ feature would indicate relatively low $O_2$ and high $O_3/O_2$ compared to modern Earth, suggesting an abiotic $O_3$ source. A mission with a very broad wavelength range could potentially detect a CO feature at 2.3 μm or 4.6 μm. This would indicate significant $CO_2$ photolysis rates as a potential abiotic source of atomic O and atmospherically derived $O_2$ and $O_3$. If the amount of CO were quantified, along with that of $O_3$ it might also indicate a significant but unutilized energy source for life. Similarly, $CO_2$ features in the infrared around 10 μm would indicate a relatively high abundance of $CO_2$ compared to modern Earth, and so a potential large source of O through $CO_2$ photolysis. These features appear in our models of high-$CO_2$ abiotic planets (Figure 5), but are absent in the spectrum of modern Earth (Figure 5, top panel) because Earth's pre-anthropogenic $CO_2$ abundance is not high enough to create these features. (The absorption feature at 9.6 μm in Earth's spectrum is caused by $O_3$, not by $CO_2$.) This strategy for discriminating false- and true-positives is similar to the one suggested by Selsis et al. (2002). In addition to the search for molecular $O_2$ features, a search for $O_2$ dimer features beyond 1 μm could also inform planet characterization (Misra et al., 2014). Although we did not explicitly model this feature, the $O_2$ concentrations in our model simulations of abiotic planets are orders of magnitude lower than those required to create such features on planets with 1 bar of total pressure.

Finally, the lack of a $CH_4$ absorption feature could indicate an abiotic source of $O_2$ or $O_3$. Looking at the entire suite of our simulations (Figure 6), there is a clear trend: decreasing $O_2$ and $O_3$ concentrations are both correlated with increasing $CH_4$ concentrations. This trend also holds for inhabited planets, but at $O_2$ and $CH_4$ concentrations that are orders of magnitude greater. Because of this trend, we expect that any atmospheres with properties beyond the parameter space explored here would exhibit undetectable amounts of $CH_4$ if they had higher $O_2$ or $O_3$ concentrations than the atmospheres simulated in this study. This stands in contrast to modern-Earth, which has orders of magnitude more $O_2$, $O_3$, and $CH_4$, and features from these gases that are correspondingly easy to detect compared to the features modeled here for abiotic planets. Although some of our abiotic simulations (e.g., the simulation of a planet around GJ876 shown in Figures 4 and 5) had detectable features from both $O_3$ and $CH_4$, placing upper limits on the column depths of these two gases would indicate potential abiotic sources, following the logic in earlier parts of this section. Such quantification – or at least the placement of low upper limits on these gases – could identify potential sources of abiotic $O_2$ and $O_3$ production.

Taken together, these features would hint at an abiotic source of the $O_3$ and would cast doubt on conclusions of a biological $O_3$ source. This indicates a need for broad-wavelength searches for life on exoplanets: to discriminate an abiotic or biological source, a UV-visible mission that attempts to observe an $O_3$ feature at 0.3 μm should at the least attempt to quantify the $O_3$ and detect $O_2$ at 0.76 μm. This means having a spectral resolution of at least R = 60 and S/N > 10 (although trades between R and S/N are possible, see Stapelfeldt et al., 2014). Preferably, it should extend into the IR so it would also be capable of observing $CH_4$ and CO absorption features. Table 4 shows that missions in the IR that observe the $O_3$ and $CH_4$ features in low resolution do not detect both for an abiotic planet; this would allow such a mission to discriminate between a biosignatures and a false positive for life. Finally, although the simulations presented here do not contain significant $SO_2$, accumulation of that gas could also lead to a high inventory of atmospheric O-atoms, or give clues to the amount of volcanic outgassing on the planet (Kaltenegger and Sasselov, 2010).

The implications of a potential false positive for life for planned or proposed exoplanet characterization missions will depend on the capability of the mission to observe the wavelengths of the features discussed above. For example, the abiotic $O_3$ absorption features are not within JWST's accessible wavelength range. JWST could observe the $O_3$ feature at 9.6 μm, but in all our simulations in which that feature was detectable, $CH_4$ concentrations were low and likely undetectable at 3.3 μm. Therefore, JWST can avoid this problem if both $CH_4$ and $O_3$ concentrations are constrained, and conclusions are not drawn on an $O_3$ absorption feature alone. Flagship-class space-based telescopes designed to characterize extrasolar planets with UV-visible-near infrared photons (e.g., Cash 2006; Soummer et al. 2009) would have the aperture size and wavelength range needed to observe the abiotic Hartley $O_3$ and $CH_4$ features. Taken on their own, the simultaneous presence of $CH_4$ and $O_3$ in these atmospheres could be misinterpreted as evidence for disequilibrium brought about by biological activity. But if the observations are further scrutinized, discrimination between biological and abiological $O_3$ sources should be possible. The large telescope size would enable the placement of tight lower limits on $O_2$ concentrations, casting doubt on the presence of biological $O_2$ production. Additionally, the observation of multiple CO absorption features would enable the identification of a potential abiotic $O_3$ source: photolysis of a large $CO_2$ reservoir. Finally, combining this information with observations of the star's energy distribution via photochemical models (such as those used here) would provide confirmation that the $O_3$ likely originated from photolysis of potentially abiotic gases.

These abiotic ozone sources would prove more difficult to missions that would attempt to use band filters to specifically target the UV absorption feature from $O_3$, such as the proposed Occulting Ozone Observatory (Pravdo et al. 2010; Savransky et al., 2010). These proposals call for an initial focus on this relatively easy to detect feature. However, this focus comes at the price of less information about the atmospheric and stellar context for the observed features. In this case, observations of the star's energy distribution would be critical. By using these data as inputs to photochemical models, the potential for an abiotic $O_3$ source could be identified. However, the biological and abiological sources for the $O_3$ feature could not be discriminated, only highlighted as possibilities. If bands targeting the smaller absorption feature from $O_2$ are also included, this could aid in $O_3$ source attribution. However, because this feature is smaller, longer observations would be required to detect it. Thus, the ideal role of this mission is likely as a "first step" in generating a prioritization list of extrasolar planets for detailed follow-up observations.

## 5. Conclusions

$O_3$ that is detectable in the UV at low spectral resolution can be caused by the presence of FUV photons that free O atoms from volcanic species such as $CO_2$ into a planetary atmosphere with high amounts of $CO_2$ and a relatively reduced chemical composition (i.e., low in fluxes of $H_2$ and other reduced gases). Because these UV fluxes and the volcanic sources of $CO_2$ and $H_2$ can be sustained for billions of years, this source of $O_3$ could be sustainable on similar timescales. This differentiates this mechanism from other abiotic sources of $O_2$ or $O_3$, such as photo-dissociation of water during in a steam-filled, runaway greenhouse atmosphere (Kasting 1997; Schindler & Kasting 2000) or models of $O_2$ or $O_3$ production that do not adequately account for redox balance in the atmosphere (Selsis et al. 2002) or the oceans (Tian et al., 2014). This mechanism is also more problematic than the "false positives" represented by the presence of $O_3$ in the atmospheres of Mars (Blamont & Chassefiere 1993; Fast et al. 2006a; Fast et al. 2006b; Fast et al. 2009; Villanueva et al. 2013) and Venus (Montmessin et al. 2011). Unlike those objects, the planet in our simulation has a mass, radius, atmospheric pressure, atmospheric opacity, and surface temperature similar to that of modern Earth. In the case of the planet around GJ 876, the atmosphere exhibits $O_3$ concentrations detectable in the UV and $CH_4$ features detectable in the VIS at low spectral resolution. These similarities make discrimination of biotic and abiotic $O_3$ more difficult that it would be for a Venus-twin or Mars-twin, as those two planets differ from Earth in identifiable planet-scale properties such as size, albedo, and atmospheric pressure and temperature structure.

Such a false positive could complicate interpretation of planetary spectra obtained by exoplanet characterization missions that operate exclusively in the UV-visible region of the spectrum. However, this chemistry only occurs on planets around stars that have high FUV fluxes, and for atmospheres that have high $CO_2$ concentrations and low $H_2$ concentrations.

Therefore, detectable $O_3$ features in the UV with low resolution can be determined to have an abiotic source if the presence of other species such as $CH_4$, CO and $O_2$ are quantified, or if attempts are made to observe multiple $O_3$ absorption features in an attempt to constrain $O_3$ concentrations. This highlights the need for contextual information and integrated models to thoroughly characterize extrasolar planets. Characterization and target star selection efforts will be greatly aided by knowledge of the stellar energy distribution incident upon the planet, and this strongly motivates future UV observations of different star types, in particular F-type stars and M dwarfs (e. g. Hawley et al. 2003; Walkowicz et al. 2008; France et al. 2012). These results also motivate the broadest possible wavelength range for spectral observations of exoplanets. The need for contextual information has been necessary when searching for life in situ (Horowitz et al. 1976; Klein et al. 1976; Navarro-González et al. 2003; Soffen 1976) on other planets, in the rock record of the Earth, and in analyses of meteorite samples; we should anticipate a similar need when we search for life across interstellar distances.

**Acknowledgments:** This manuscript was jointly led by Shawn D. Domagal-Goldman (SDDG) and Antígona Segura (AS), who both independently discovered the main phenomenon described in the manuscript. Victoria S. Meadows (VSM) and Tyler D. Robinson (TDR) provided expertise with respect to planetary spectra and a modern Earth spectrum for comparison. Mark W. Claire provided photochemical modeling advice and key inputs into the manuscript. All authors contributed to multiple revisions of the text.

Amit Misra provided us with critical advice on calculating S/N ratios. James F. Kasting, Kevin France, Michael H. New, Feng Tian, Kevin Zahnle, the Virtual Planetary Laboratory NAI team, and an anonymous reviewer provided helpful comments on prior versions of this manuscript or on the figures presented herein.

This work was performed as part of the NASA Astrobiology Institute's Virtual Planetary Laboratory, supported by the National Aeronautics and Space Administration through the NASA Astrobiology Institute under solicitation No. NNH05ZDA001C. SDDG, MWC, and TDR acknowledge additional support from the Oak Ridge Associated Universities NASA Postdoctoral Program; SDDG did much of the work on this project while an NPP Management fellow in residence at NASA Headquarters, MWC while in residence at the University of Washington and TDR while in residence at NASA Ames.

**Figures and Tables:**

**Figure 1**. A) UV Stellar energy distributions (SED's) for the σ Boötis (F2V), the Sun (G2V), ε Eridani (K2V), AD Leonis (M3.5V), and GJ 876 (M4V) for a planet receiving the integrated energy Earth receives from the Sun (1360 W $m^{-2}$), with a slight correction applied to account for how the albedo of a planet will change around different star types (after Segura et al., 2005). B) Absorption cross-sections for $CO_2$, $O_2$ and $O_3$, corresponding to reactions (R1), (R2), and (R4), respectively. The two panels are on the same scale, allowing estimates of the relative rates of these photolysis reactions expected around the stars studied here.

**Figure 2.** Color contours of A) $O_2$ and B) $O_3$ optical depth of atmospheres of abiotic planets with different compositions and stellar fluxes. Each contour shows the $O_3$ color contours (colors in log-space) as a function of the $CO_2$ concentration at the surface and the $H_2$ volcanic outgassing rate. Contours are shown for each permutation of stellar fluxes for five different stars – σ Boötis

(F2V), the Sun (G2V), ε Eridani (K2V), AD Leonis (M3.5V), and GJ 876 (M4V) – and for three different constraints on the ocean redox balance.

**Figure 3**. A) Modeled reflectance spectra, from 100-400 nm (0.1-0.4 μm), for the photochemical simulations of abiotic planets around all five stars in this study (labeled colored lines), compared to the reflectance spectrum of modern Earth (black line). B-F) Reflectance spectra (black lines) for our model planets around σ Boötis (F2V), the Sun (G2V), ε Eridani (K2V), AD Leonis (M3.5V), and GJ 876 (M4V), respectively, where we have individually removed the spectral influences of $O_3$, $O_2$, $CH_4$, CO, and $CO_2$ from the model (blue, green, red, goldenrod, and grey lines, respectively). The absorption features of $O_2$, $O_3$, $CH_4$, and CO are labeled. Unlabeled features are from $CO_2$ or $H_2O$.

**Figure 4.** A) Modeled spectra, from 0.5-5 μm, for the photochemical simulations of abiotic planets around all five stars in this study (labeled colored lines), compared to the spectrum of modern Earth (black line). B-F) Spectra (black lines) for our model planets around σ Boötis (F2V), the Sun (G2V), ε Eridani (K2V), AD Leonis (M3.5V), and GJ 876 (M4V), respectively, where we have individually removed the spectral influences of $O_3$, $O_2$, $CH_4$, CO, and $CO_2$ from the model (blue, green, red, goldenrod, and grey lines, respectively). The absorption features of $O_2$, $O_3$, $CH_4$, and CO are labeled. Unlabeled features are from $CO_2$ or $H_2O$. All spectra in this figure have been convolved to a resolution of 70.

**Figure 5.** A) Modeled spectra, from 5-20 μm, for the photochemical simulations of abiotic planets around all five stars in this study (labeled colored lines), compared to the spectrum of modern Earth (black line). B-F) Spectra (black lines) for our model planets around σ Boötis (F2V), the Sun (G2V), ε Eridani (K2V), AD Leonis (M3.5V), and GJ 876 (M4V), respectively, where we have individually removed the spectral influences of $O_3$, $O_2$, $CH_4$, CO, and $CO_2$ from the model (blue, green, red, goldenrod, and grey lines, respectively). The absorption features of $O_2$, $O_3$, $CH_4$, and CO are labeled. Unlabeled features are from $CO_2$ or $H_2O$. All spectra in this figure have been convolved to a resolution of 70.

**Figure 6.** $O_2$ (teal circles) and $O_3$ (purple triangles) column depths from all our simulations, plotted as a function of the $CH_4$ column depth. $CH_4$ is used a proxy for the amount of reducing gases in the atmosphere because unlike other more measures (such as $H_2$ column density), it has discernable absorption features in terrestrial planet atmospheres.

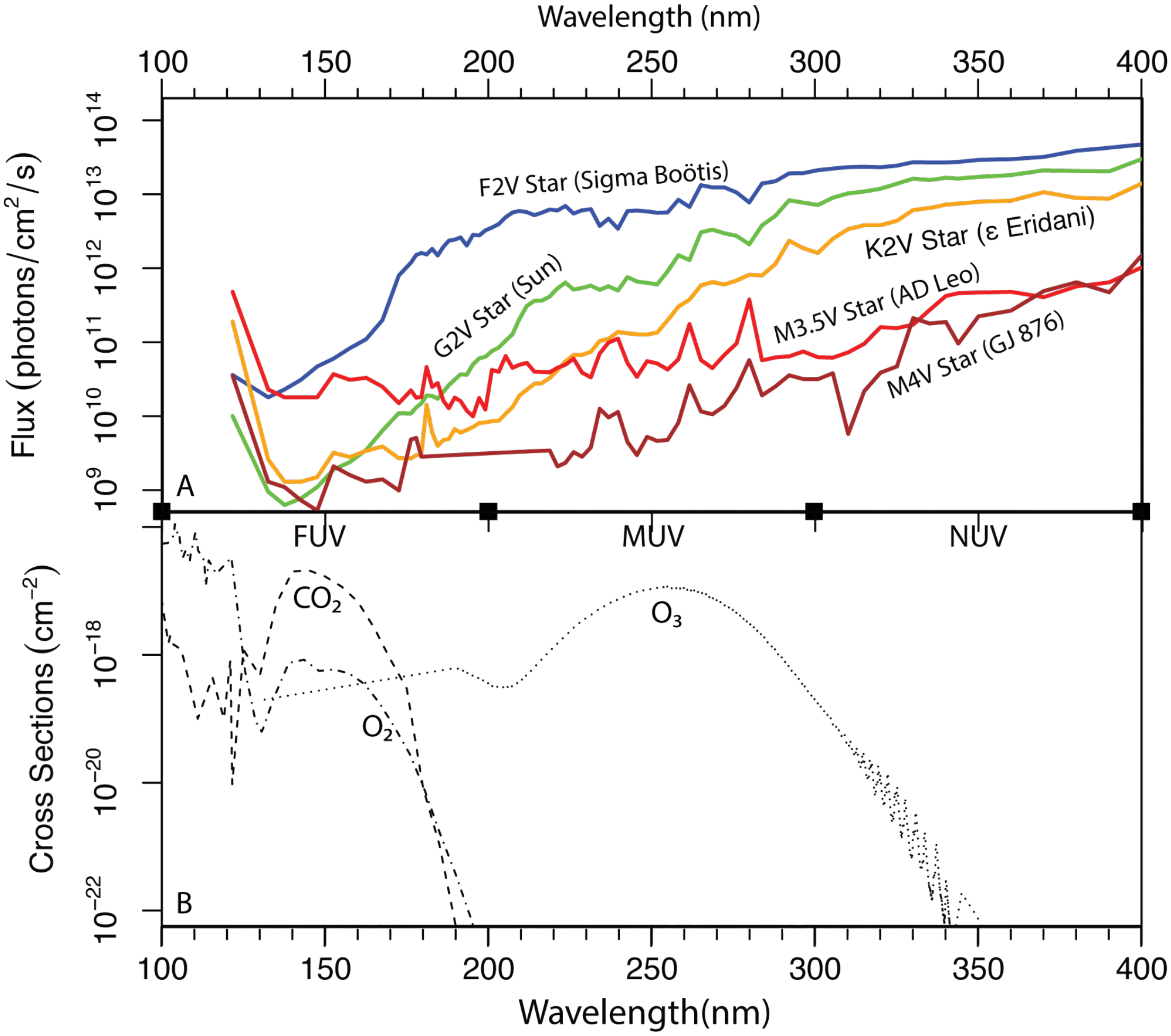

**Figure 1**. A) UV Stellar energy distributions (SED's) for the σ Boötis (F2V), the Sun (G2V), ε Eridani (K2V), AD Leonis (M3.5V), and GJ 876 (M4V) for a planet receiving the integrated energy Earth receives from the Sun (1360 W $m^{-2}$), with a slight correction applied to account for how the albedo of a planet will change around different star types (after Segura et al., 2005). B) Absorption cross-sections for $CO_2$, $O_2$ and $O_3$, corresponding to reactions (R1), (R2), and (R4), respectively. The two panels are on the same scale, allowing estimates of the relative rates of these photolysis reactions expected around the stars studied here.

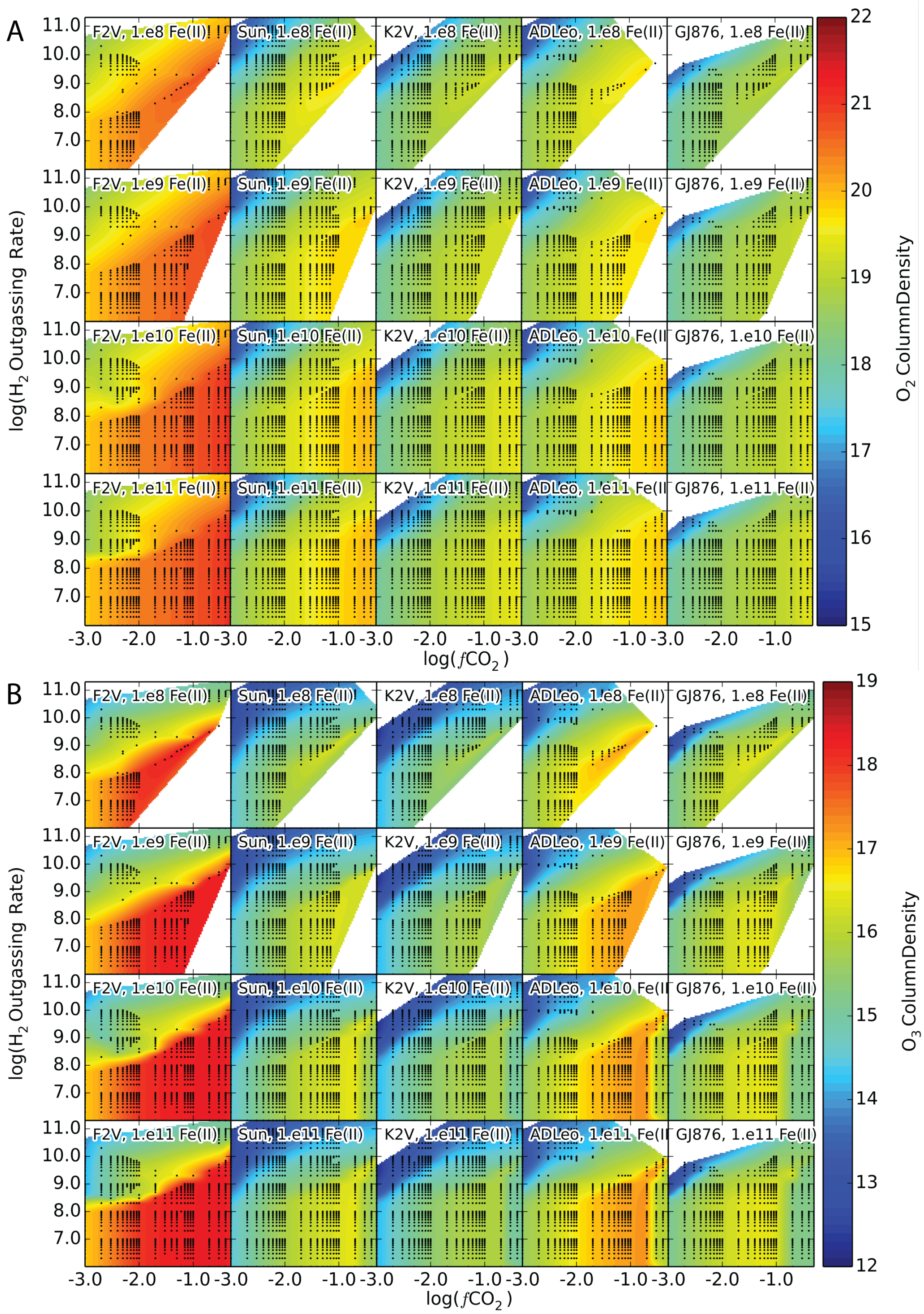

**Figure 2.** Color contours of A) $O_2$ and B) $O_3$ optical depth of atmospheres of abiotic planets with different compositions and stellar fluxes. Each contour shows the $O_3$ color contours (colors in log-space) as a function of the $CO_2$ concentration at the surface and the $H_2$ volcanic outgassing rate. Contours are shown for each permutation of stellar fluxes for five different stars – σ Boötis (F2V), the Sun (G2V), ε Eridani (K2V), AD Leonis (M3.5V), and GJ 876 (M4V) – and for three different constraints on the ocean redox balance.

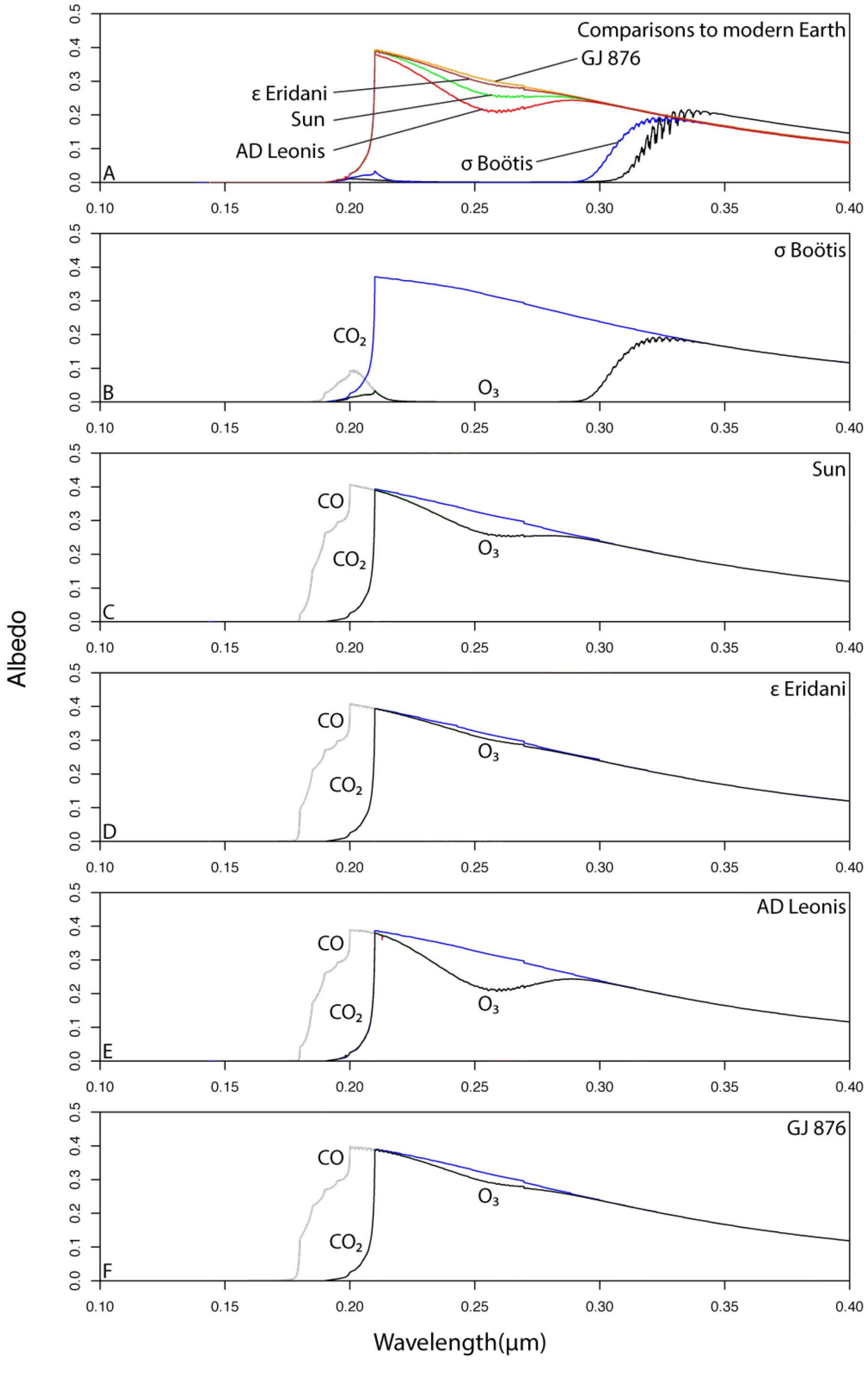

**Figure 3**. A) Modeled reflectance spectra, from 100-400 nm (0.1-0.4 μm), for the photochemical simulations of abiotic planets around all five stars in this study (labeled colored lines), compared to the reflectance spectrum of modern Earth (black line). B-F) Reflectance spectra (black lines) for our model planets around σ Boötis (F2V), the Sun (G2V), ε Eridani (K2V), AD Leonis (M3.5V), and GJ 876 (M4V), respectively, where we have individually removed the spectral influences of $O_3$, $O_2$, $CH_4$, CO, $CO_2$, and $H_2O$ from the model (blue, green, red, goldenrod, grey, and cyan lines, respectively).

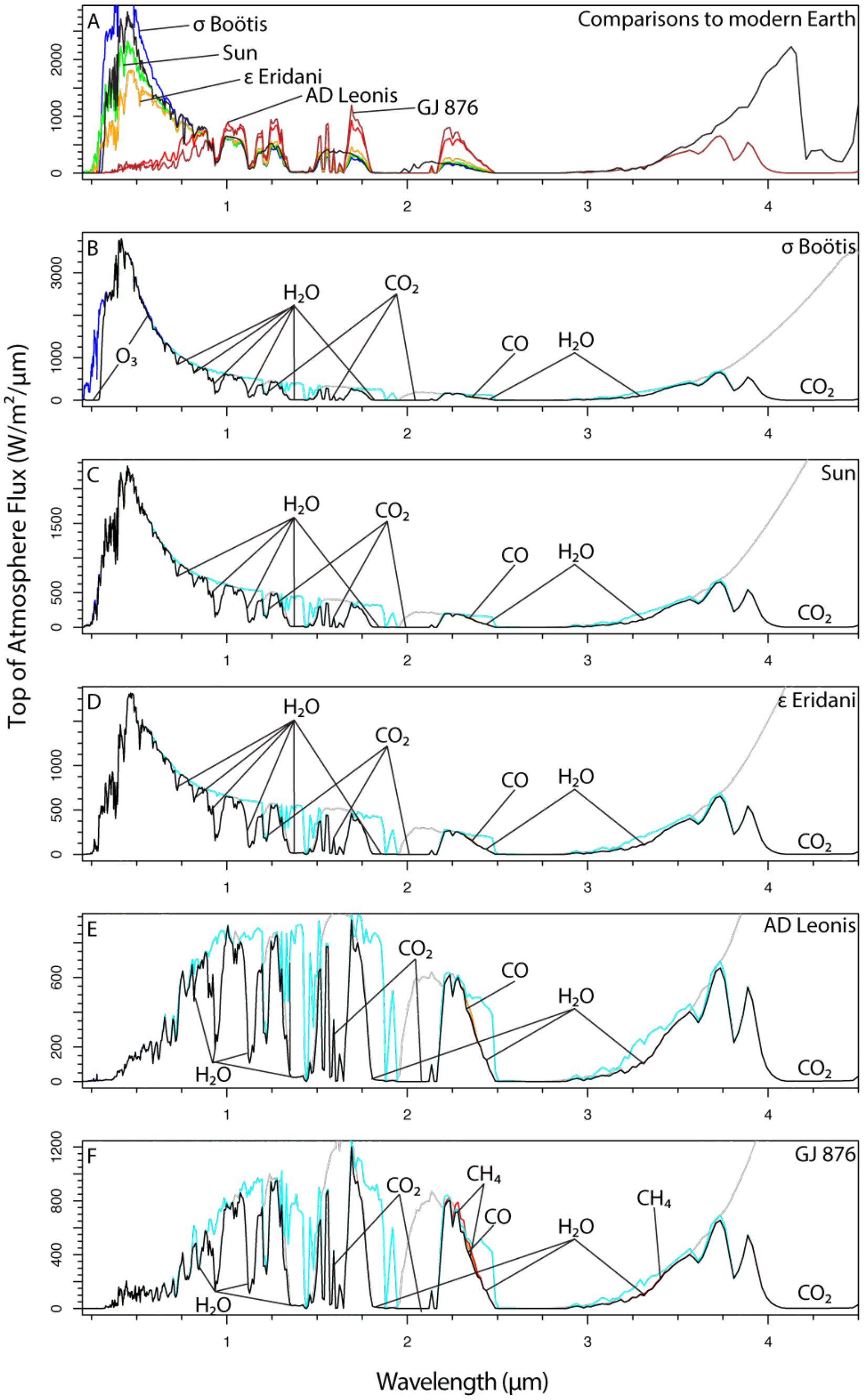

**Figure 4.** A) Modeled spectra, from 0.2-4.5 μm, for the photochemical simulations of abiotic planets around all five stars in this study (labeled colored lines), compared to the spectrum of modern Earth (black line). B-F) Spectra (black lines) for our model planets around σ Boötis (F2V), the Sun (G2V), ε Eridani (K2V), AD Leonis (M3.5V), and GJ 876 (M4V), respectively, where we have individually removed the spectral influences of $O_3$, $O_2$, $CH_4$, CO, $CO_2$, and $H_2O$ from the model (blue, green, red, goldenrod, grey, and cyan lines, respectively). The $CH_4$ feature in the GJ876 spectrum is only quantifiable once the overlapping $H_2O$ and CO features can be accounted for with independent constraints on the concentrations of these gases.

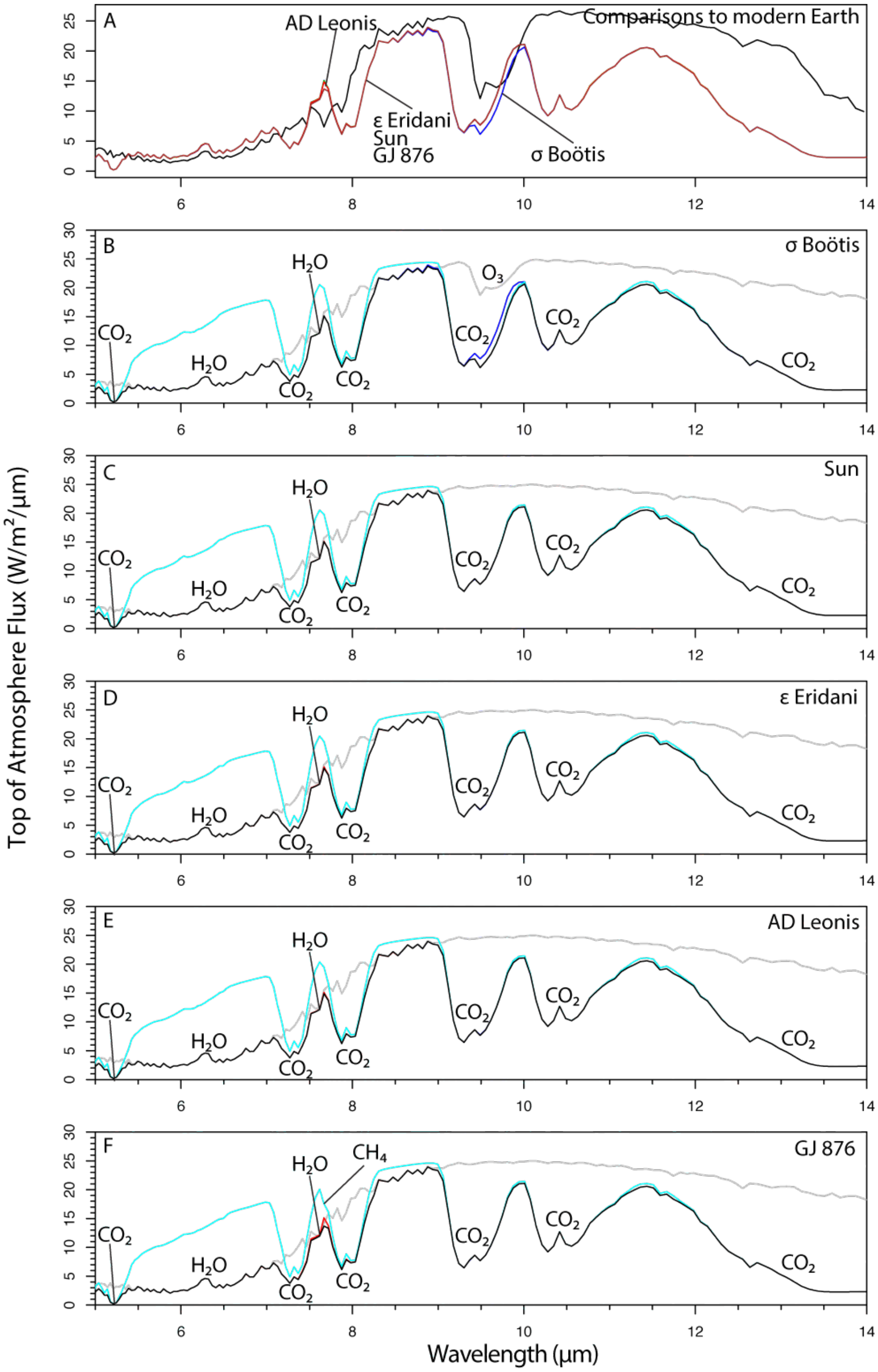

**Figure 5.** A) Modeled spectra, from 5-14 μm, for the photochemical simulations of abiotic planets around all five stars in this study (labeled colored lines), compared to the spectrum of modern Earth (black line). B-F) Spectra (black lines) for our model planets around σ Boötis (F2V), the Sun (G2V), ε Eridani (K2V), AD Leonis (M3.5V), and GJ 876 (M4V), respectively, where we have individually removed the spectral influences of $O_3$, $O_2$, $CH_4$, CO, $CO_2$, and $H_2O$ from the model (blue, green, red, goldenrod, grey, and cyan lines, respectively). The $O_3$ feature in the σ Boötis spectrum and the $CH_4$ feature in the GJ 876 spectrum can only be confidently identified once overlapping features from $CO_2$ and $H_2O$ (respectively) are removed.

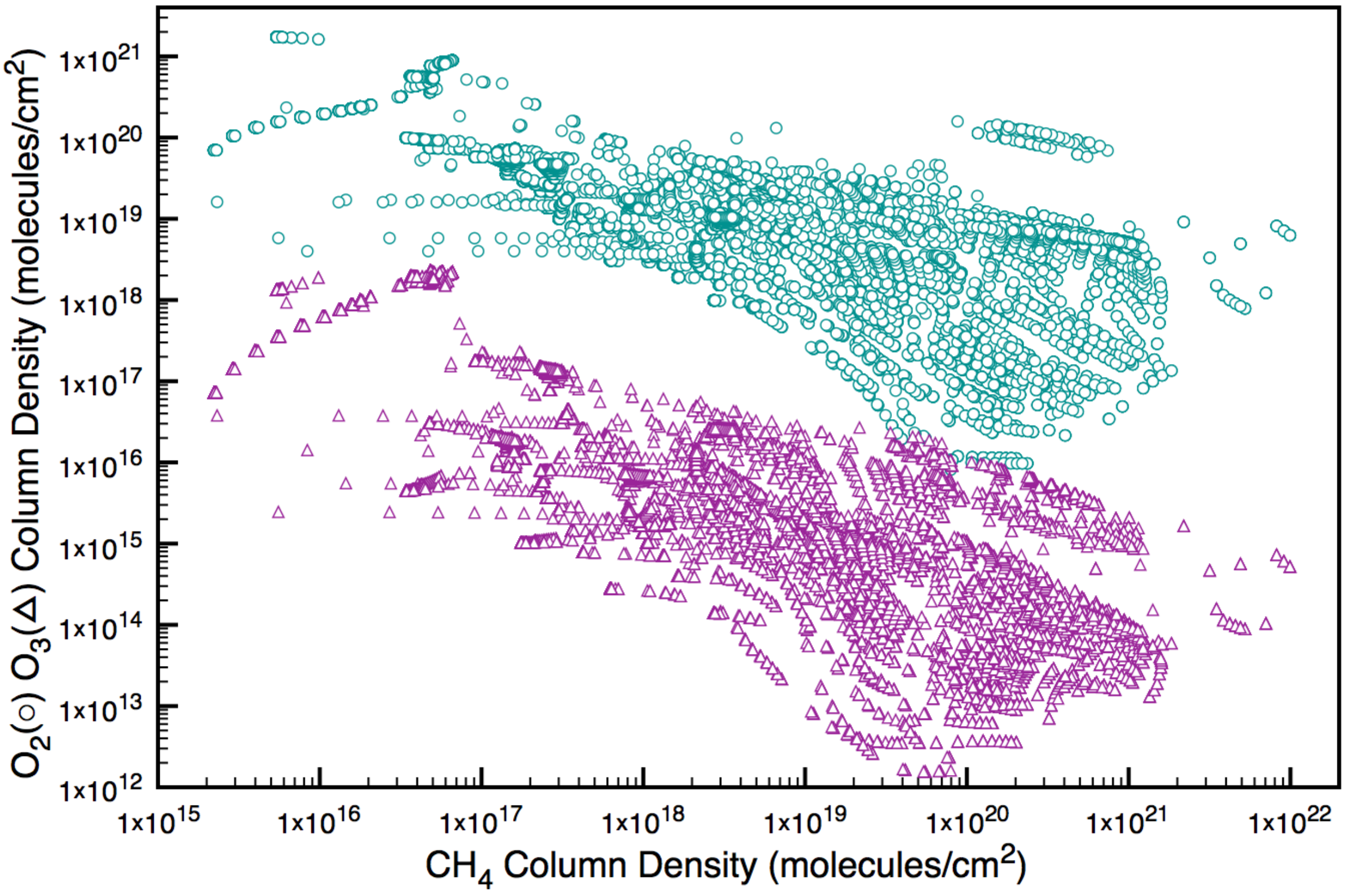

**Figure 6.** $O_2$ (teal circles) and $O_3$ (purple triangles) column depths from all our simulations, plotted as a function of the $CH_4$ column depth. $CH_4$ is used a proxy for the amount of reducing gases in the atmosphere because unlike other more measures (such as $H_2$ column density), it has discernable absorption features in terrestrial planet atmospheres. Biogenic atmospheres exhibit the same trends, but do so at higher $CH_4$, $O_2$, and $O_3$ concentrations. In other words, modern Earth (and similar planets) should plot further up and to the right of this diagram.